# Direct observation of phase change accommodating hydrogen uptake in bimetallic nanoparticles

*Lívia P. Matte,[1,2] Maximilian Jaugstetter,[3] Alisson S. Thill,[1] Tara P. Mishra,[4] Carlos Escudero,[5] Giuseppina Conti,[2,6] Fernanda Poletto,[7] Slavomir Nemsak,[2,6*] Fabiano Bernardi[1*]*

[1] Programa de Pós-Graduação em Física, Instituto de Física, Universidade Federal do Rio Grande do Sul (UFRGS); Porto Alegre, 91501-970, Brazil.

[2] Advanced Light Source, Lawrence Berkeley National Laboratory; Berkeley, 94720, USA.

[3] Chemical Sciences Division, Lawrence Berkeley National Laboratory; Berkeley, 94720, USA.

[4] Materials Science Division, Lawrence Berkeley National Laboratory; Berkeley, 94720, USA.

[5] ALBA Synchrotron Light Source, Cerdanyola del Vallès; Barcelona, 08290, Spain.

[6] Department of Physics and Astronomy, University of California; Davis, 95616, USA.

[7] Departamento de Química Orgânica, Instituto de Química, Universidade Federal do Rio Grande do Sul (UFRGS); Porto Alegre, 91501-970, Brazil.

ABSTRACT

Hydrogen holds great promise as a cleaner alternative to fossil fuels, but its efficient and affordable storage remains a significant challenge. Bimetallic systems, such as Pd-Ni, present a promising option for storing hydrogen. In this study, using the combination of different cutting-edge X-ray and electron techniques, we observed the transformations of Pd-Ni nanoparticles, which initially consist of a NiO-rich shell surrounding a Pd-rich core but undergo a major transformation when interacting with hydrogen. During the hydrogen exposure, the Pd core breaks into smaller pockets, dramatically increasing its surface area and enhancing the hydrogen storage capacity, especially in nanoparticles with lower Pd content. The findings provide deep understanding of the morphological changes at the atomic level during hydrogen storage and contribute for designing cost-effective hydrogen storage using multi-metallic systems.

The use of non-renewable energy sources is directly related to climate change due to the emission of greenhouse gases. It is estimated that natural gas and oil reserves will be depleted before 2070,[1] creating energy insecurity in economies based on these fuels. Thus, it is urgent to dramatically change the world's energy matrix to a more renewable and sustainable one. Hydrogen is one of the most promising candidates to replace fossil fuels since it is available from renewable energy sources.[2] It presents a gravimetric density 3 times bigger than gasoline, for example.[2] Another benefit is that byproducts of energy generation using hydrogen consist mainly of $H_2O$. However, the main challenge for the large-scale use of hydrogen as an energy carrier in light-duty vehicles is the hydrogen storage process. The currently available methods make it commercially disadvantageous. Hence, the US Department of Energy (DOE) set audacious targets for hydrogen storage materials that should be fulfilled by 2025, but we are still far away from reaching it.[3]

A promising way to store hydrogen is using solid materials,[4] where the hydrogen can be either adsorbed at the surface or stored with the formation of a hydride phase. For instance, hydrogen can be adsorbed on the surface of activated carbon, which has a high gravimetric capacity, but presents a very low volumetric capacity.[4] On the other hand, complex metal hydrides, such as $LiAlH_4$, present a high gravimetric and volumetric capacity, but too high desorption temperatures.[4] Thus, the discovery of improved storage materials is still needed before a full-scale commercialization can be realized.

The material used to adsorb hydrogen should present an adsorption energy in the quasi-molecular bonding regime, between roughly 0.2 eV (19.3 kJ/mol $H_2$) and 0.6 eV (57.9 kJ/mol $H_2$).[5] The major candidates are metallic nanostructures, where the adsorption energy can be adjusted by their morphology.[2] Pd is known to have a high affinity with hydrogen and to form a stable Pd-H bond.[6] However, the current Pd nanostructures do not get even close to fulfilling the DOE targets.[6] Nevertheless, the adsorption energy may be adjusted also through the stoichiometry in bimetallic nanoparticles.[4] Switching from monometallic to bimetallic nanostructures allows the creation of new atomic sites for hydrogen adsorption, thus improving the hydrogen storage capacity.[7] Previously, we demonstrated that NiO nanofoams present a quasi-molecular bonding with hydrogen thereby presenting a lower hydrogen affinity compared to Pd.[8] Therefore, a mixture of NiO to Pd in the form of Pd-NiO nanoparticles shows great promise for hydrogen storage applications. The inclusion of NiO is helpful also to reduce the cost compared to monometallic Pd. Previous studies demonstrated that the bimetallic Pd-Ni nanoparticles are great candidates for promoting efficient hydrogen storage.[9–11] Thus, it is fundamental to thoroughly understand the atomic events occurring during hydrogen storage in these bimetallic nanoparticles to design improved hydrogen storage systems.

We herein present an approach utilizing complementary *in-situ* techniques to determine atomic- and nano-scale transformations during the hydrogen storage process in $Pd_xNi_{100-x}$ (x = 90, 75, 50, and 25) nanoparticles. Each technique is dedicated to probing different chemical and structural properties of the nanoparticles during the process. X-ray absorption spectroscopy (XAS) was used to observe the changes in the local atomic order around Ni and Pd atoms, X-ray photoelectron spectroscopy (XPS) probes the chemistry of components, and grazing incident X-ray scattering (GIXS) observed structural transformations. Measurements are complemented by electron energy loss spectroscopy (EELS) for nano-scale insights into the particles' morphology and chemistry. The combination of all techniques reveals complex atomic mechanisms promoting the hydrogen storage in these bimetallic nanoparticles. Furthermore, our results present a new cost-effective solution for hydrogen storage using $Pd_xNi_{100-x}$ nanoparticles.

RESULTS/DISCUSSION

We performed *ex-situ* characterization of morphology and composition of nanoparticles using high-angle annular dark field - scanning transmission electron microscopy (HAADF-STEM) imaging. Figure 1(a) presents a typical image obtained for $Pd_{25}Ni_{75}$ nanoparticles, showing a nanoparticle with a diameter of around 20 nm, which is the mean value obtained for all samples from the TEM analysis, as presented in the histogram of size distribution (Figure S1). Figure 1(b) presents the XRD measurements, the indexing with the crystal phases, and the Rietveld refinement. NiO and PdO phases are present in all samples, besides the presence of metallic Pd (Pd(0)) phase in smaller amounts in the Pd-richer samples, which agrees with XANES measurements in Figure S2. In general, NiO and PdO phases have similar crystallite size for all the samples (see Table S1). The XRD and TEM results show that the bimetallic nanoparticles exhibit similar diameter, morphology, and oxidation state between the samples. Moreover, the Pd content obtained from surface-sensitive XPS measurements in the as-prepared state is always smaller than that obtained from EDS measurements (Figure S3), indicating that all as-prepared samples present a core-shell-like structure with a Ni-rich shell and a Pd-rich core region. In addition, EDS maps show that Ni and Pd are well dispersed in the samples (Figure S4).

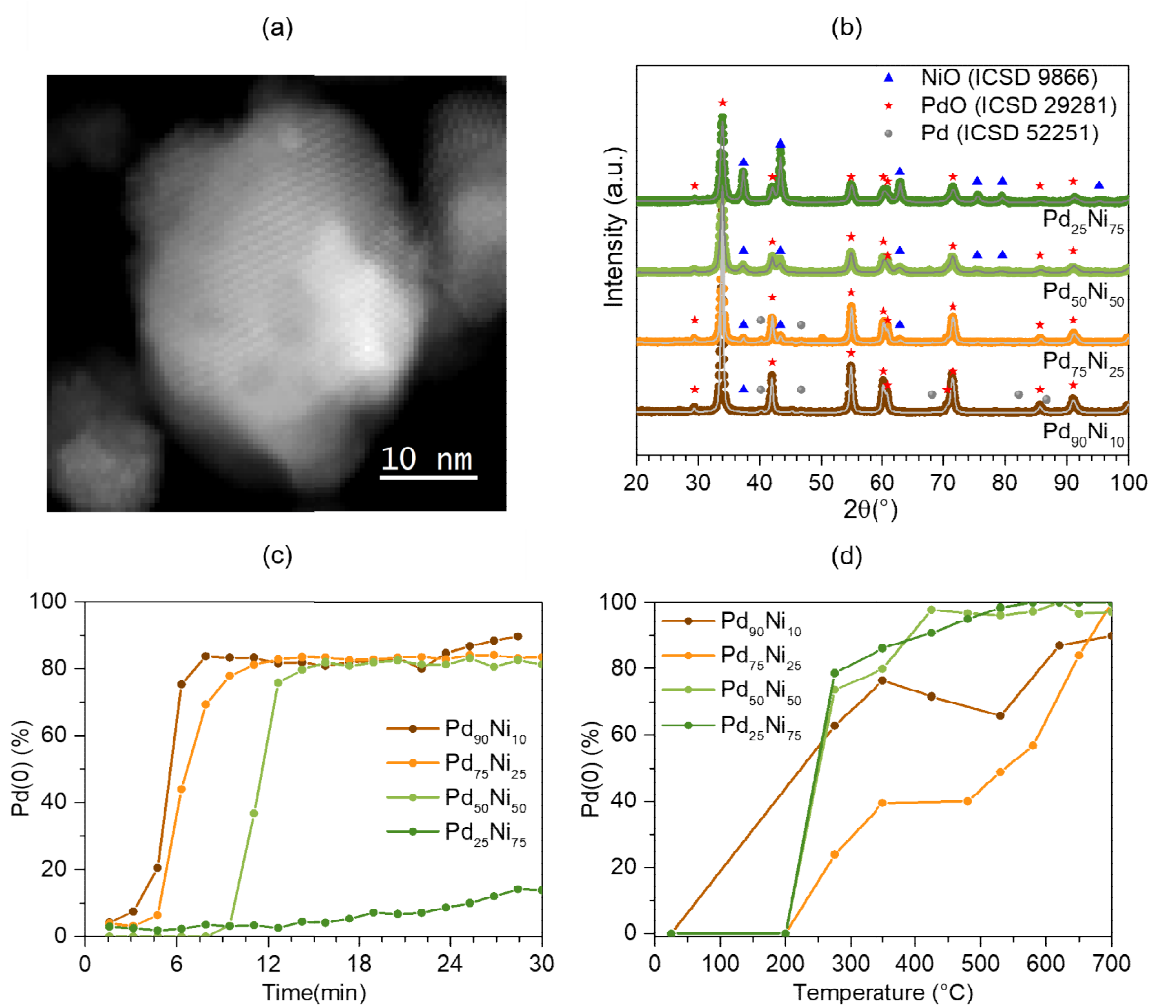

.

**Figure 1.** (a) Typical HAADF-STEM image of the $Pd_{25}Ni_{75}$ sample. (b) XRD measurements of the $Pd_xNi_{100-x}$ samples. The points represent the experimental data, and the gray line represents the fit obtained from the Rietveld refinement. The blue triangle, red star, and gray circle show the position of the Bragg reflections related to the NiO, PdO, and Pd(0) crystalline phases, respectively. (c) Pd(0) fraction as a function of time obtained from in-situ XANES measurements during the 30 mL/min 4% $H_2$ + 96% He exposure at room temperature and (d) on the nanoparticles' surface obtained from the XPS measurements under UHV as a function of temperature.

The reduction of Pd atoms is an essential step in the hydrogen storage process since Pd(0) has a much higher hydrogen storage capacity than PdO.[12] The opposite occurs with Ni atoms, where NiO presents adsorption energy in the quasi-molecular bonding regime, while metallic Ni (Ni(0)) presents adsorption energy in the chemisorption regime.[8] Figure 1(c) shows the Pd(0) content obtained from the *in-situ* XANES analysis as a function of time during the hydrogen exposure at RT (Figure S5). In line with previous reports, the initial high amount of PdO quickly reduces to Pd(0).[13,14] Nevertheless, Pd is not fully reduced at the end of process, in accordance to the literature that shows that the O diffusion process is too slow to fully reduce PdO to metallic Pd at room temperature.[15] However, the time needed to reduce nanoparticles from PdO to Pd(0) strongly depends on the Pd/Ni ratio. The higher the amount of Pd, the faster the Pd reduction at RT under $H_2$ exposure, suggesting that Ni plays an inhibitory role in the reduction process of Pd.

The near-surface region of the as-prepared samples exhibits only NiO and PdO chemical components, as observed by XPS (Figure S7). Figure 1(d) shows the atomic content of reduced

Pd(0) as a function of temperature, obtained from the analysis of the AP-XPS spectra (Figure S8) measured under ultra-high vacuum (UHV). It is observed that the total surface reduction occurs at higher temperatures for samples containing higher amounts of Pd. This is in contrast to what is observed under $H_2$ atmosphere.

Aiming to observe the changes in the local atomic order around Ni and Pd atoms during the hydrogen storage process, we performed *in-situ* EXAFS measurements. Figure 2 shows the Fourier Transform (FT) of the EXAFS oscillations (shown in Figure S9 and S10) for nanoparticles with different Pd/Ni ratios during $H_2$ exposure at RT and atmospheric pressure. The FT data of all the samples are very similar to Pd(0) and NiO standards. However, the $Pd_{25}Ni_{75}$ sample also shows a small contribution from the Pd-O scattering path along with the Pd(0). Surprisingly, there is a clear decrease in the intensity of the peak related to the Pd-Pd scattering path in the nanoparticles in comparison to Pd(0) standard, but this reduction in intensity does not occur for Ni-O and Ni-Ni paths. This is related to a reduced size or higher temperature of the sample. Since all measurements were conducted at RT, it shows some atomic rearrangement around Pd atoms different than Ni atoms. Under $H_2$ exposure, the Pd-Pd scattering peak shifts to higher R values as compared to the Pd(0) standard, indicating lattice expansion upon $H_2$ uptake. Furthermore, no shifts in the Ni-Ni or Ni-O scattering paths are observed compared to NiO standard during $H_2$ exposure.

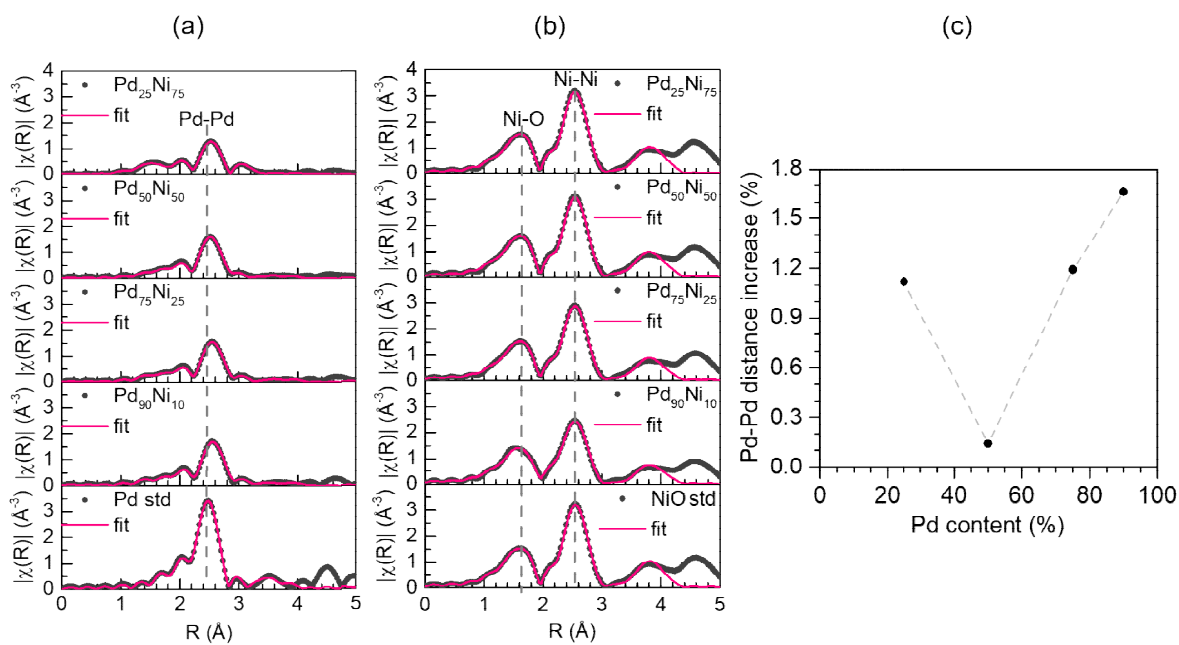

**Figure 2.** Fourier Transform of the EXAFS oscillations at the (a) Pd K edge and (b) Ni K edge during $H_2$ exposure at RT and atmospheric pressure. The black dots represent the data measured and the pink line the fit performed. (c) Increase of the Pd-Pd atomic distance in comparison to Pd standard position as a function of Pd content in the nanoparticles.

Figure 2(c) displays the relative increase in the Pd-Pd scattering path distance from the coordination shell when compared to the Pd(0) reference (see Table S2 for the fitting parameters). There is an increasing trend of the Pd-Pd interatomic distance (up to 1.7% increase) with the Pd amount, except for $Pd_{50}Ni_{50}$ nanoparticles that present only a slight increase. However, no shifts are observed in the Ni-Ni or Ni-O scattering peaks (Table S3).

To observe the initial stages of the hydrogen storage process starting at the surface of the nanoparticles, we performed AP-XPS measurements. The initial heat treatment at 250 °C under ultra high vacuum (UHV) facilitates organic remnants removal and the reduction of PdO to Pd(0), as demonstrated in Figure 1(d). After the annealing treatment, the samples were exposed to 0.1 mbar $H_2$ while AP-XPS measurements were performed. Figure 3 (a) and (b) show typical Pd 3d and Ni 3p AP-XPS measurements for the $Pd_{25}Ni_{75}$ nanoparticles using incident beam photon energies of 695 eV and 1000 eV, respectively. Different photon energies allow probing different depths in the nanoparticles since the inelastic mean free path (λ) of photoelectrons coming from the Pd 3d electronic region is around 13 Å (1000 eV) or 8 Å (695 eV). The spectra for all other bimetallic samples are shown in Figure S11 and S12. We carried out the measurements at RT, before and during exposure to 0.1 mbar of $H_2$. The peak at around 336 eV (Figure 3 (a)) is assigned to the Pd(0) component.[16] For $Pd_{90}Ni_{10}$ nanoparticles, we observe that the nanoparticles are not fully reduced after thermal treatment since they show a shoulder at higher binding energies (around 338 eV), which is associated with PdO (see Figure S11 (a)).[12] This shoulder disappears when the sample is exposed to $H_2$, thus showing the full reduction of PdO to Pd(0). The small binding energy shifts in Pd 3d and Ni 3p regions are discussed in more detail in Note S1. It is important to note that the absolute binding energy values and shifts for suspended nanoparticles are not always indicative of a chemical change. The electrostatic alignment between the particle and substrate can result in apparent shifts,[17] as it is believed to be the case here as well. The presence of a doublet at around 349 eV is also observed, which is attributed to the Ca 2p core level that comes from the synthesis procedure. The main component at the Ni 3p AP-XPS region (Figure 3(a)) at around 70 eV comes from NiO.[18] In this energy region, the peak at around 65 eV is associated with Na 2s electronic region, which similarly to Ca, is a contaminant coming from the synthesis and transfer of nanoparticles. However, no contamination was observed by EDS in these samples (see Figure S13), showing that this contamination may be in small amounts on the surface of the nanoparticles.

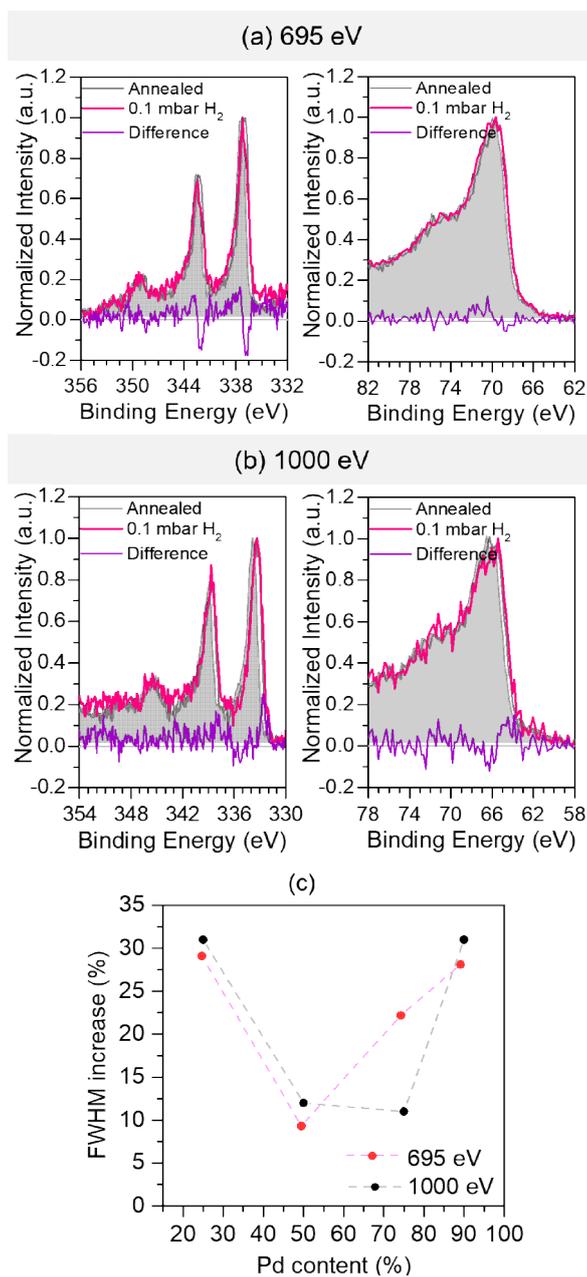

**Figure 3.** Typical AP-XPS measurements of the $Pd_{25}Ni_{75}$ nanoparticles in the Pd 3d and Ni 3p energy regions using a photon energy of (a) 695 eV and (b) 1000 eV. The grey and pink lines

represent the measurement after annealing under UHV and during 0.1 mbar $H_2$ exposure for 2h at RT, respectively. The difference between the spectrum during the $H_2$ exposure process and the spectrum measured after annealing is presented below each spectrum in purple. (c) Relative increase in the FWHM of the peak related to Pd(0) when comparing the Pd 3d spectra after annealing and during $H_2$ exposure, as a function of Pd content in the nanoparticles.

The introduction of $H_2$ atmosphere induces an increase in the FWHM value of the Pd(0) component. The widening starts to occur at $H_2$ pressures as low as $1\times10^{-7}$ mbar but stabilizes at 0.1 mbar, as shown in Figure S14. Figure 3(c) displays the relative increase in the FWHM value of the Pd(0) component during $H_2$ exposure in comparison to the value before the insertion of $H_2$ atmosphere obtained from the fitting procedure performed, as shown in Figure S15 (see Table S4). All the samples present a clear increase in the FWHM value in this period for both photon energies used. This is indicative of the presence of a second Pd component related to the Pd-H bonding. However, due to the binding energy proximity of Pd(0) and Pd-H components, which is around 0.2 eV,[19] it is not possible to distinguish the two components in the AP-XPS measurements. In almost all the samples, this increase in the FWHM value is more significant for the measurements performed using an incident beam of 1000 eV than 695 eV. The more significant increase in the FWHM value for the higher excitation energy of 1000 eV indicates the presence of more Pd-H bonds in the subsurface layers. In general, higher the amount of Pd in the composition, higher the relative increase in the FWHM value. However, $Pd_{25}Ni_{75}$ sample also presents a high relative increase due to the relatively low amount of Pd atoms that can readily form Pd-H bonds.

On the other hand, Ni 3p AP-XPS region typically shows no changes in intensity or shape due to H$_2$ exposure, thus indicating the preferential bonding of hydrogen with Pd atoms, in agreement with *in-situ* XAS results. However, for the highest Ni amount sample (Figure 3 (a,b)), a small increase in the FWHM of the Ni 3p region is also observed. This may indicate that hydrogen starts to be co-adsorbed on Ni atoms in the close proximity of Pd, which readily forms Pd-H bonds. However, this is hard to detect through the *in-situ* XAS measurements presented because they are bulk-sensitive.

To confirm that the changes observed by *in-situ* XAS and AP-XPS measurements are indeed due to the hydrogen storage capacity of these samples, we performed hydrogen storage measurements. For these measurements, 1 wt.% of Pd$_x$Ni$_{100-x}$ nanoparticles were supported over activated carbon. The measurements were performed after exposing the samples to 1 atm H$_2$ atmosphere at RT for 6 hours. Figure S16 shows the hydrogen uptake as a function of the Pd contend in the sample and Table S5 shows the gravimetric and volumetric capacity for each sample. It is observed that the hydrogen uptake follows the same trend as the increase of the FHWM value of the Pd peak obtained from AP-XPS measurements and the increase of the Pd-Pd distance obtained from in-situ XAS measurements. Thus, the changes observed in AP-XPS and *in-situ* XAS measurements can be attributed to the hydrogen storage effect. Furthermore, the gravimetric capacity value is reproduced at least after 1 cycle of hydrogen release and hydrogen exposure again.

Since hydrogen storage occurs predominantly on the Pd atoms, it is essential to evaluate how the Pd/Ni ratio at the surface changes during the hydrogen storage process. Thus, the Pd atomic fraction at the sample surface was calculated by normalizing the Pd 4p and Ni 3p areas by their respective photoionization cross-sections.[20] Inelastic mean free path of these core-electrons was around 15 Å (1000 eV) and 11 Å (695 eV). Figure S3 shows that in all cases, Pd content at the surface is significantly smaller than that obtained by EDS. As explained above, it indicates the presence of a core-shell-like structure with a Ni-rich shell and a Pd-rich core region, even during the $H_2$ exposure. Also, there is a slight increase in the Pd content at the surface region during hydrogen exposure, showing a change in the atomic arrangement of the nanoparticles.

We performed AP-GIXS measurements to gain a further understanding of the atomic rearrangement taking place in the nanoparticles during the treatment. Figure 4(a) presents a typical AP-GIXS measurement for the as-prepared $Pd_{25}Ni_{75}$ sample. To minimize the influence of particle clustering and island formation, we integrated over the in-plane scattering vector $q_{xy}$ in the region indicated by the red outline in Figure 4(a) to obtain the intensity as a function of the out-of-plane scattering vector $I(q_z)$ (Figure S20).[21] Figure 4(b) presents the pair distance distribution function (PDDF) obtained from the inverse Fourier transform of the line-cut. A decrease in the maximum R value, which defines the maximum distance between two scattering atoms in a single nanoparticle, is observed after the annealing process. Similarly, there is a clear decrease in the Radius of Gyration (Figure 4(c)), which is a shape-independent indicator of particle size,[22,23] with the treatment employed, mainly after heating to 250 °C. These results indicate the contraction of the $Pd_{25}Ni_{75}$ nanoparticles after the annealing process. This contraction can be attributed to the reduction of PdO to Pd(0).

Figure 4 (d-f) presents the electron density reconstruction of the PDDF of the $Pd_{25}Ni_{75}$ nanoparticles under the different treatments applied (see Note S18).[24] Pd(0) has the highest electron density of all compounds in the nanoparticles, while NiO has the lowest electron density value (Table S6). Figure 4(d) shows that the as-prepared $Pd_{25}Ni_{75}$ nanoparticles have an ellipsoid shape of ~30 nm length and ~20 nm width, which is in good agreement with the obtained AFM images of these particles (see Figure S17) and conducted GIXS simulations (see Figure S18). The particles have high electron density in the core region, while a lower density is observed in the shell region, indicating the presence of a Pd-rich core and Ni-rich shell. The size and shape of as-prepared nanoparticles are confirmed by HAADF-STEM (Figure 1(a)). Figure 4(e) presents the PDDF reconstruction of the $Pd_{25}Ni_{75}$ nanoparticles after annealing in UHV. A shape change from an ellipsoid-like to a spherical-like particle is clearly observed. Again, a higher electron density is present in the core region of these nanoparticles as compared to the shell. However, this high-density core is more diffuse in comparison to the as-prepared case. Lastly, the $Pd_{25}Ni_{75}$ nanoparticles during exposure to 0.1 mbar $H_2$ retain their spherical shape (Figure 4(f)). Surprisingly, the inner structure of the nanoparticle undergoes a transformation, in which segmentation of high-electron density pockets inside the nanoparticle and their diffusion towards the surface is observed during hydrogen exposition.

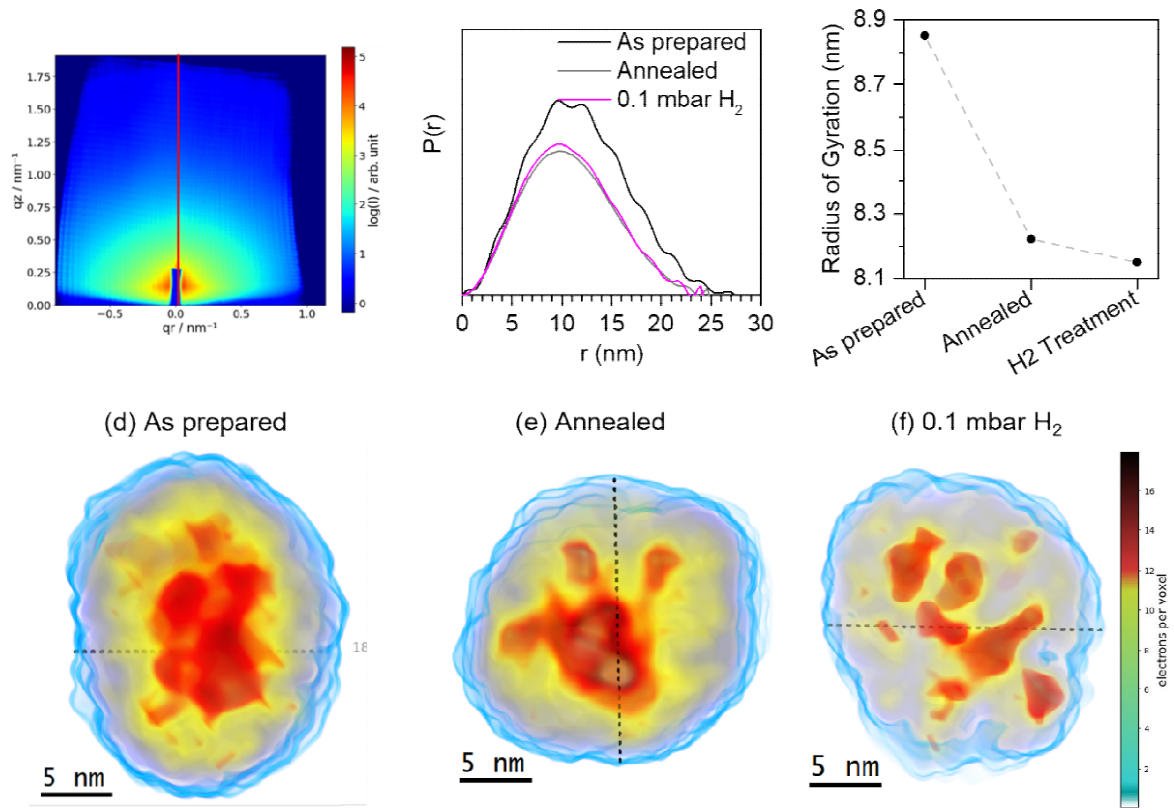

**Figure 4.** (a) Typical AP-GIXS measurement of the as-prepared $Pd_{25}Ni_{75}$ nanoparticles. The red straight line represents the position of the cut used in the analysis. (b) Pair distance distribution function (PDDF) obtained from the inverse Fourier transform of the cut presented in (A) for the $Pd_{25}Ni_{75}$ as prepared, annealed, and under 0.1 mbar $H_2$ atmosphere, presented in black, gray, and magenta, respectively. (c) Radius of gyration as a function of the treatment applied. On the bottom, reconstruction of the electronic density inside the $Pd_{25}Ni_{75}$ nanoparticles for the (d) as prepared, (e) annealed, and (f) under 0.1 mbar $H_2$ atmosphere. In red is presented the region with higher electronic density, yellow, medium electronic density, and blue, lower electronic density.

The same analysis procedure was applied to the other $Pd_xNi_{100-x}$ compositions, as shown in Figure 5(a-c) for 0.1 mbar $H_2$ exposure. The same qualitative behavior is observed for all three stoichiometries (see Figure S19 for the annealed sample), *i.e.*, the core region with a higher electron density in comparison to the shell region and the formation of fragmented high electronic density pockets. It explains the small intensity at FT for Pd K edge and the same intensity for Ni K edge in comparison to Pd(0) and NiO standards (Figure 2).

Figure 5(d) shows the relative expansion of the nanoparticles during the exposure to 0.1 mbar $H_2$ in comparison to before $H_2$ introduction as a function of Pd content. A contraction of the $Pd_{25}Ni_{75}$ and $Pd_{50}Ni_{50}$ nanoparticles is observed, while the $Pd_{75}Ni_{25}$ nanoparticles undergo an expansion during $H_2$ exposure. The expansion is related to the increase in the Pd lattice size when hydrogen is adsorbed in the interstitial site, as determined by *in-situ* XAS measurements. However, due to the low amount of Pd and the complicated rearrangement of the Pd pockets inside the nanoparticles, $Pd_{25}Ni_{75}$ and $Pd_{50}Ni_{50}$ nanoparticles exhibit a size contraction.

Moreover, Figure 5(e) shows the relative expansion of the sphere enclosing all the high-electron density pockets as a function of the Pd content in the sample. Such analysis utilizing an equivalent sphere is typically used for fractal nanoparticle systems to compare different shape configurations.[25] For lower Pd concentrations, a pocket enclosure expansion of 20% is observed. This occurs by the dissolution of high electron density regions inside the nanoparticles with lower Pd content. A contraction occurs for the nanoparticles containing a higher Pd concentration. Figure 5(f) shows the increase in the FWHM of the electronic density distribution curve across the particle, as shown in Figure S20.

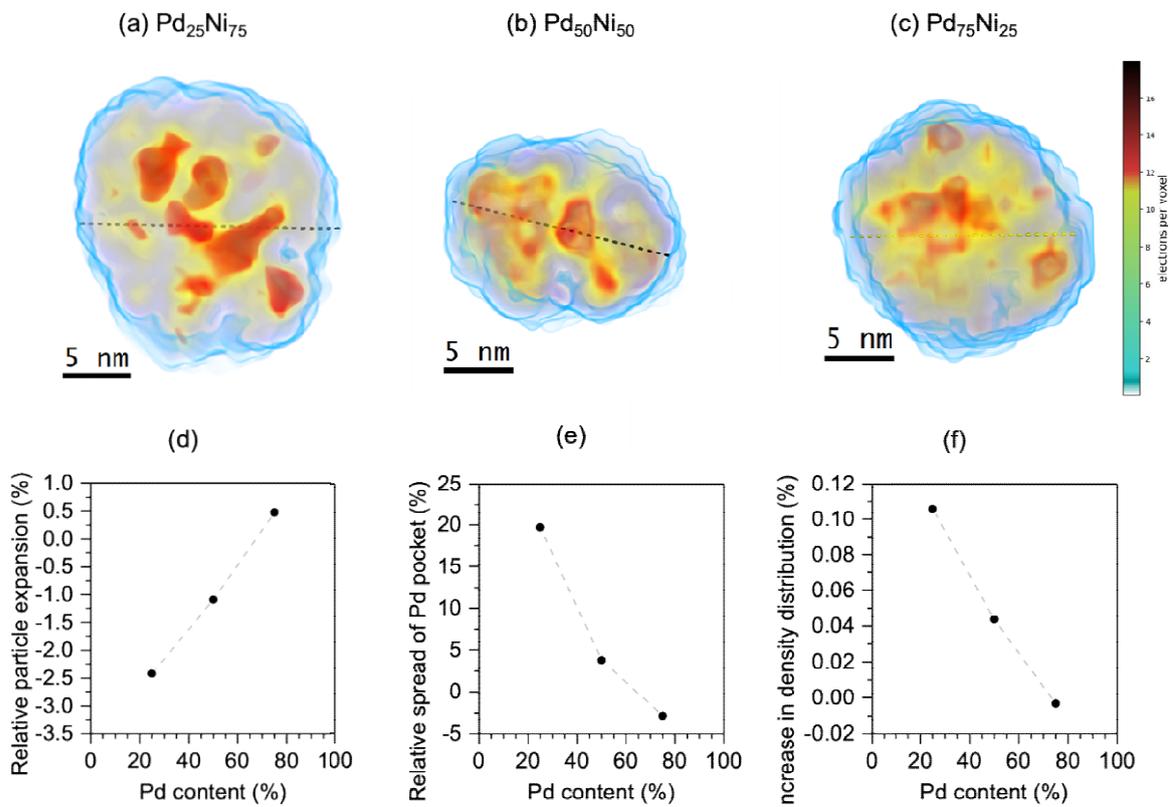

**Figure 5.** Reconstruction of the electronic density under 0.1 mbar $H_2$ atmosphere for the (a) $Pd_{25}Ni_{75}$, (b) $Pd_{50}Ni_{50}$ and (c) $Pd_{75}Ni_{25}$. In red is presented the region with higher electronic density, yellow, medium electronic density, and blue, lower electronic density. (d) Relative increase in particle mean diameter, calculated from the smallest particle enclosing sphere to account for the changes in shape and surface roughness, during $H_2$ treatment. (e) Relative expansion of the fractal enclosing sphere used to define the outer boundaries of the high density Pd core before and during $H_2$ treatment in percent. (f) Relative increase in the FWHM value of the curve relative to the electronic density distribution inside the nanoparticles.

To validate the conclusions obtained from AP-GIXS analysis, STEM-EELS measurements were performed. Figure 6 shows a typical STEM-HAADF image of the as-prepared $Pd_{25}Ni_{75}$ nanoparticles with around ~ 20 nm diameter and its Pd and Ni compositional maps (see Figure S21 for a typical EELS spectrum). The Ni composition maps (Figure 6(b)) clearly show pockets of Ni-deficient regions in the nanoparticles. These deficient regions are also observed in the O composition map obtained by integrating the O K edge due to the atomic density difference of PdO and NiO (shown in Figure S22). The Pd compositional map (Figure 6(c)) shows the presence of Pd-rich clusters of around 3 to 10 nm size located in the Ni deficient regions. A composite map formed by overlaying the Ni (green) and Pd (magenta) compositional maps is shown in Figure 6(d). It is evident that multiple pockets of Pd-rich clusters are embedded in the NiO matrix of a single nanoparticle, as validated with several other nanoparticles (Figure S23).

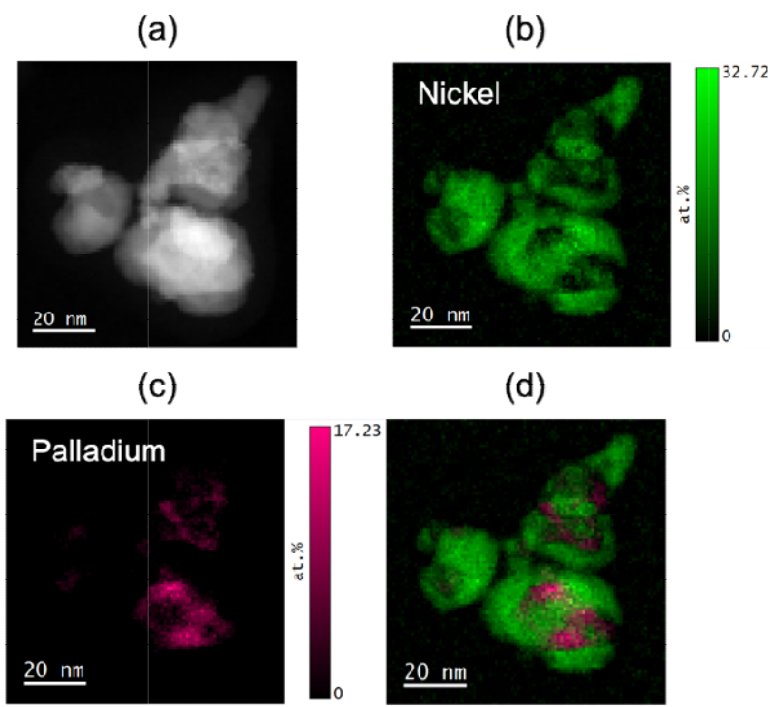

**Figure 6.** (a) STEM-HAADF image of a typical as prepared Pd25Ni75 nanoparticle from which (b) Ni compositional map was obtained by integration of L edge and (c) Pd compositional map was obtained by integration of the M edge. (d) A composite map overlapping the Ni (green) and Pd (magenta) maps. The color bars in these images show the atomic percentage which is calculated based on all the available elements in the spectral window from 230 eV to 1419.3 eV, which includes both C K-edge (~284 eV) and O K-edge (~532 eV).

All the as-prepared $Pd_xNi_{100-x}$ nanoparticles exhibit an internal structure with a Pd-rich core and Ni-rich shell, as indicated by the compositional difference between bulk-sensitive EDS and surface-sensitive XPS analysis. This phase separation was confirmed by AP-GIXS (Figure 4(d)) and EELS measurements (Figure 6). In addition, it was observed that the as-prepared samples present mainly NiO and PdO compounds. For the metallic Pd-Ni system, the opposite is theoretically expected, *i.e.*, a Pd-rich shell and a Ni-rich core,[26] but the Ni-rich surface comes due to the stability of NiO. Upon annealing under UHV or exposure to $H_2$ at RT, Pd in these nanoparticles undergoes reduction from PdO to Pd(0). On the other hand, from the AP-XPS and *in-situ* EXAFS measurements (Figure 2 and 3) it is evident that Ni remains as NiO, which agrees with previous reports.[27] In addition, during reduction under $H_2$ atmosphere at RT, we observed that the sample with higher amount of Ni present a slower reduction than the sample with higher amount of Pd. This can be related to the existing core-shell-like structure. The Ni-richer samples present a thicker (and consequently more impermeable) shell, acting like a diffusion barrier for the $H_2$ molecules. It makes the reduction kinetics of PdO slower due to the slower diffusion of the $H_2$ molecules through the Ni-shell, which is a necessary step to reach the Pd-rich core. The opposite is observed during the reduction process under UHV. In this case, since the thermally-induced reduction process does not depend on the diffusion of $H_2$ molecules through the NiO layer, the Ni-rich sample reduces in lower temperature than the Pd-rich sample.

During exposition to $H_2$ atmosphere, the EXAFS measurements showed an increase in the Pd-Pd interatomic distance (Figure 2 (c)), while no difference is observed in the Ni K-edge. From the d-band theory,[28] it is known that the strain influences the hydrogen storage properties. Lattice expansion upshifts the d-band center, leading to a stronger adsorbate binding. The increase in the Pd-Pd interatomic distance may also be related to the storage of hydrogen in the interstitial sites around Pd atoms, thus increasing the lattice size, as observed in the literature.[29,30] Similarly, we observed an increase in the FWHM of the Pd(0) (Figure 3 (c)) peak at Pd 3d AP-XPS measurements under $H_2$ atmosphere. This increase in the FWHM is also attributed to the storage of H atoms on Pd, since, as explained before, it indicates the presence of a second Pd component (Pd-H). In both cases, no change is observed in the Ni K-edge or Ni 3p peak spectra. Thus, it shows that the hydrogen atoms get exclusively adsorbed on the Pd atoms. In addition, it is observed that the highest increase in the R value and FWHM increase the highest Pd content on the studied stoichiometries. Thus, it is expected that $Pd_{90}Ni_{10}$ sample would have the highest hydrogen storage capacity, which was confirmed by hydrogen storage measurements. Intuitively, although $Pd_{75}Ni_{25}$ and $Pd_{25}Ni_{75}$ present a similar increase in the R value and FWHM, the former is expected to exhibit a higher hydrogen storage capacity due to the higher Pd content. However, we observed a similar gravimetric capacity for these two samples. This may be related to a higher propensity for phase separation of the Pd as observed by AP-GIXS measurements. Therefore, both the Pd-Pd distance measurements (Figure 2(c)), relative FWHM increase of the Pd(0) peak at Pd 3d spectra (Figure 3(c)) and gravimetric capacity show a relative minimum at the $Pd_{50}Ni_{50}$ composition.

Moreover, it was previously observed that 1wt.% of Pd nanoparticles supported on activated carbon present an storage capacity around 0.7 wt.%.[31] In the current study, we show that the addition of only 25% of Pd gives the same hydrogen storage capacity. The hydrogen storage capacity of the activated carbon alone was also measured, and it is around 0.08 wt.%. Thus, the bimetallic nanoparticles (with exception of $Pd_{50}Ni_{50}$) present an increase in the gravimetric capacity by more than a factor of 8. On the other hand, $Pd_{50}Ni_{50}$ nanoparticles show a reduced hydrogen storage of 0.4 wt%, which is in line with previous findings.[11] Since Pd is a much more expensive resource compared to Ni, it is important to highlight the economic importance of these results. Adding a small amount of Pd (~ 25%) is enough to have a hydrogen uptake value similar to the monometallic Pd nanoparticles. It indicates that such high Ni content nanoparticles can provide an economically viable solution for hydrogen storage.

The phase segregation observed by AP-GIXS (see Figure 4) is induced by the storage of hydrogen mainly onto the Pd atoms. Interchange of atoms in a core-shell structure under a reducing or oxidizing atmosphere has been widely observed for many different bimetallic systems at elevated temperatures.[32] However, the interchange of atoms from the Pd core into the nanoparticle shell at RT, as observed here, causes the core to split into Pd pockets inside a Ni shell. In addition, a formation of well-defined high-density structures is observed. To the best of our knowledge, this is the first time the segregation of the core of the nanoparticle into small pockets has been observed *in-situ* induced by exposure to hydrogen at room temperature.

In the samples with higher Pd content, the Pd is more dispersed inside the nanoparticles, as observed in the AP-GIXS measurements (Figure S17). This might be either due to the lack of NiO in the nanoparticles to form a proper core-shell structure or to the formation of a solid solution. Theoretical ternary Ni-Pd-O convex hull drawn using the Materials Project (Figure

S22) does show $Pd_{50}Ni_{50}$ (~ 43 meV) close to the stable convex hull.[33,34] Previous work has shown that such structures might get stabilized at higher temperatures.[35,36] This stabilization is due to entropic effects which depend critically on how the target alloy composition of interest responds to the temperature effects in contrast to all relevant competing phases.[37,38] However, to the best of our knowledge, experimental Ni-Pd-O phase diagrams have not been reported till date.

The effect of the phase separation can be visualized by the reconstructed electronic density obtained from the AP-GIXS measurements (Figure 5 (a) and Figure S17). If the $Pd_{50}Ni_{50}$ composition is stabilized and forms a solid solution, the $Pd_{75}Ni_{25}$ composition would phase separate into pure Pd rich and $Pd_{50}Ni_{50}$ regions, thereby explaining the presence of very high Pd-rich regions. However, for both $Pd_{50}Ni_{50}$ and $Pd_{25}Ni_{75}$ samples, no Pd rich regions are observed. This is most likely because these compositions are either stable (the case of $Pd_{50}Ni_{50}$) or the phase separate in the stable $Pd_{50}Ni_{50}$ and Ni-rich compositions (the case for $Pd_{25}Ni_{75}$).

From the AP-GIXS measurements, it is evident that during the hydrogen storage process, considerable rearrangement of Pd atoms is observed. In general, upon $H_2$ storage, the increase in particle size is in proportion to the Pd content of the nanoparticle. However, the relative Pd pocket expansion and relative distribution compared to the annealed nanoparticles decreases with the increase in the Pd content under the $H_2$ atmosphere. This is most likely due to the lack of a well-delineated Pd core in high Pd-containing samples. However, in all cases the hydrogen storage leads to a rearrangement of Pd, forming Pd-rich pockets.

Previously, evidence of the atomic rearrangement during hydrogen storage process was also observed in a Pd-rich core and Pt-rich shell nanoparticles. Tayal *et al.*[39] concluded that the storage of hydrogen in this sample transformed the core-shell structure into a mixed Pd-Pt alloy. This process increased the Pd-Pt interfacial area, where the hydrogen was adsorbed, then increasing the hydrogen storage capacity of this sample. Similarly, in this work the sample with the smallest amount of Pd presents many small Pd pockets that are very well delimited and segregated from the NiO phase. This process increases the interfacial area between the Pd and NiO phases where the hydrogen can be stored.[7,39,40]

In fact, a deeper look at the Ni K edge XANES spectra (Figure S23) shows changes in the edge position. In all samples, the introduction of $H_2$ induces a shift to smaller binding energies, as observed elsewhere.[39] It indicates that NiO is not just a bystander during the hydrogen storage process, but the hydrogen is adsorbed at the interface between the NiO shell and Pd pockets. It compensates for the small amount of Pd in these nanoparticles, leading to strong changes in the parameters related to hydrogen storage from AP-XPS measurements. On the other hand, in Pd-rich nanoparticles, aside from starting to form Pd pockets, this structure is not clearly defined, presenting a diffused transition between Pd pockets and NiO shell. These results indicate that the hydrogen storage process occurs efficiently in these Pd pockets, where a plum pudding-like system is formed. It also reveals the underlying mechanisms, which can be used to reduce the use of noble metals while retaining the $H_2$ storage capacity in bimetallic nanoparticles. This could be also a possible explanation for similar gravimetric capacity observed for the $Pd_{25}Ni_{75}$ and $Pd_{75}Ni_{25}$ samples. Considering this, the main mechanism of hydrogen storage in the Pd-Ni nanoparticles consists on the hydrogen absorption at the interface of Pd pockets and NiO medium.

CONCLUSIONS

In this work, we explored the behavior of Pd$_x$Ni$_{100-x}$ core-shell nanoparticles during hydrogen storage, revealing intriguing transformations in both their surface and internal structure. By employing advanced in-situ X-ray techniques, we gained a comprehensive understanding of how these nanoparticles change morphologically and chemically at the atomic level. While hydrogen primarily interacts with Pd, NiO plays a crucial role by acting as a buffer that allows Pd to form internal pockets. As hydrogen is adsorbed, these pockets fragment, significantly increasing their surface area available for hydrogen storage, even if the overall particle size decreases. Notably, samples with lower Pd content exhibit hydrogen uptake similar to pure Pd, showcasing that a small addition of Pd (~25%) to the NiO system presents a great potential to reduce costs in hydrogen storage technologies while maintaining or increasing the performance. Our insights also suggest that future designs of bimetallic nanoparticles should focus on enhancing the formation of hydrogen-active nano-pockets of material - paving the way for more efficient and cost-effective hydrogen storage solutions. Deeper understanding of hydrogen storage mechanisms, as presented here, will drive the development of sustainable energy technologies necessary for transformation of energy industry.

METHODS/EXPERIMENTAL

*Synthesis*

We synthesized Pd$_x$Ni$_{100-x}$ nanoparticles with different compositions, namely Pd$_{90}$Ni$_{10}$, Pd$_{75}$Ni$_{25}$, Pd$_{50}$Ni$_{50}$, and Pd$_{25}$Ni$_{75}$. The total mass of nickel chloride hexahydrate plus palladium acetate was kept constant at 92.9 mg in all cases, and its individual values were chosen to achieve the desired Pd/Ni atomic ratios. Initially, nickel chloride hexahydrate (NiCl$_2$ • 6H$_2$O) and glucose (0.039g) were totally dissolved in MilliQ® water (7.2 mL), while palladium acetate (Pd(OCOCH$_3$)$_2$) was dissolved in monoolein (1-Oleoyl-rac-glycerol technical grade, ~ 40% purity) (20 g). This monoolein phase was kept in an ultrasonic bath for 5 minutes while it was constantly stirred with a spatula. The aqueous phase was added to the monoolein phase and mixed for 1 minute using a spatula. This mixture rested for 10 minutes. After this, it was added to a water bath at 80 °C for 1 hour and mixed with a spatula every 5 minutes. After the bath, the mixture rested for 2 h, and then they were calcined under air at 500 °C for 4 h, forming a fine dark powder containing the nanostrucures.

### *Transmission Electrons Microscopy (TEM)*

We performed TEM measurements at CM-UFMG using a Tecnai G2 Spirit Biotwin microscope operated at 120 kV. For the measurements, the samples' powder was dispersed in MilliQ® water and left in the ultrasound bath for 15-20 min. Then, three drops of this solution were added to a carbon-coated Cu grid. The size distribution histograms of the Pd$_x$Ni$_{100-x}$ nanoparticles were obtained using ImageJ (Version 6.0) software. The diameter of each nanoparticle was calculated from the total area of its projection in the TEM screen, considering spherical nanoparticles.

### *Electron Energy Loss Spectroscopy (EELS)*

We performed EELS measurements using the TEAM 1 microscope (double aberration corrected Thermo Fisher Scientific Titan 80-300) at the National Center for Electron Microscopy (NCEM). The EELS dataset was collected at 300 kV using a Gatan Continuum spectrometer, a convergence angle of approximately 30 mrad, and a collection angle of 110 mrad. The width of the zero-loss peak was measured to be 1.2 eV, and a 0.35 eV/channel dispersion was used to collect the spectra. The composition map was obtained by integrating the Ni L edge ($L_2$ ~ 872 eV and $L_3$ ~ 855 eV), Pd M edge ($M_4$ ~ 340 eV and $M_5$ ~ 335 eV) and O K edge at 532 eV.

*X-ray Diffraction (XRD)*

We performed the XRD measurements in a Rigaku Ultima IV diffractometer at CNANO-UFRGS using Cu Kα (1.5406 Å) radiation with a graphite monochromator and working at 40 kV and 17 mA. $Pd_xNi_{100-x}$ samples were fixed in a glass blade for the XRD measurements. The data were collected for a 2θ range between 20° and 100°, using a scanning step of 0.05°. Pattern indexing was performed using reference standards obtained from the ICSD database.[41] The Rietveld refinement was performed on FullProof[42] software (version July-2017). The Instrumental Resolution Function (IRF) was obtained from a $SiO_2$ standard. Rietveld refinements were carried out using the pseudo-Voigt profile function of Thompson, Cox, and Hastings[43] and a linear background. During the refinement, the Debye-Waller overall factor and the atom occupancy were fixed to the crystallographic information file (cif) value obtained from the literature.

*In situ X-ray Absorption Spectroscopy (XAS)*

We performed in situ XAS measurements at NOTOS beamline from ALBA Synchrotron Light Source in transmission mode at the Pd K edge (24350 eV) and Ni K edge (8333 eV). For the measurements, the nanoparticles powder was supported over activated carbon (Activated Charcoal Norit®) with 7 wt% using an ultrasound bath for 20 min. For the Pd K edge and Ni K edge measurements, 13 mm pellets were produced with 75 mg of $Pd_xNi_{100-x}$/C and 25 mg of BN. For the measurements, the samples were initially exposed to 30 ml/min $N_2$ and heated to 150 °C where they remained for 15 min. Then, the nanoparticles were cooled down to room temperature (RT) under 30 mL/min He. At RT, the samples were exposed to 30 mL/min of 4% $H_2$ + 96% He and remained in this condition until its complete stabilization, i.e., until no more changes in the X-Ray Absorption Near Edge Structure (XANES) spectra were observed. The XAS measurements were performed during the full treatment employed. The XANES and EXAFS measurements for Ni K edge were performed integrating 0.24 s/point and with a 0.93 eV and 1 eV energy step, respectively, while for the Pd K edge the integration time was 0.2 s/point and the energy step used was 0.5 eV and 1 eV, respectively. The Extended X-Ray Absorption Fine Structure (EXAFS) spectra were analyzed in accordance with the standard procedure of data reduction[44] using IFEFFIT.[45] The EXAFS signal $\chi(k)$ was extracted, and then Fourier transformed using a $k^2$-weighted $\chi(k)$ and a Kaiser-Bessel window with a k-range of 10.0 Å$^{-1}$ for both edges. The phase shift and amplitudes were obtained with the FEFF6 code by using a fcc metallic Pd, tetragonal PdO, fcc metallic Ni and a fcc NiO cluster with 6 Å radius each. For the fit of the EXAFS oscillations, the amplitude reduction factor ($S0_2$) was fixed at 0.85, as obtained from the fit of both Pd(0) and Ni(0) standards. The R-factor obtained from the analysis was always lower than 0.01, which demonstrates the excellent agreement between the proposed model and the experimental result.

***Ambient Pressure X-ray Photoelectron Spectroscopy (AP-XPS) and Ambient Pressure Grazing Incidence X-ray Scattering (AP-GIXS)***

We performed AP-XPS and AP-GIXS measurements at APPEXS endstation at 11.0.2 beamline of ALS Synchrotron.[46] For the measurements, the nanoparticles powder was dispersed in ethanol and added to an ultrasonic bath for 10 min. After that, a silicon wafer was covered 3 times with the solution containing the nanoparticles, thus forming a thin layer. After that, the samples were inserted into an ambient pressure reaction chamber, where they were heated to 250 ºC under vacuum to clean the sample's surface and to reduce from PdO to Pd(0). Then, the samples were cooled to room temperature, and once at this temperature, they were exposed to 0.1 mbar $H_2$. AP-GIXS measurements were performed for the samples as prepared, after annealing, and during hydrogen exposure. AP-XPS measurements were performed before and after every AP-GIXS measurement. For the AP-XPS measurements, a SPECS PHOIBOS 150 NAP analyzer was used, with a photon beam energy of 1000 eV and 695 eV. Energy regions collected were survey, Pd 3d, Ni 3p, Pd 4p, Si 2p, C1s, and O1s. A step of 1 eV and 0.1 eV, dwell time of 0.2 s, and pass energy of 10 eV were used for the survey and high-resolution spectra, respectively.

KOLXPD software (v. 1.8.0) was used for the XPS analysis. For the fitting of the high-resolution spectra, a Shirley-type background was used. The fitting procedure, shown in Figure S15, was performed by fixing the distance between Pd(0) and PdO and Pd(0) satellite at 1.3 eV[47] and 6.7 eV,[48] respectively. An asymmetric Doniach Sunjic peak was used for the fit of the Pd(0) component, and Voigt peaks were used for all the other chemical components. The Gaussian width of the peaks was fixed at the same value for all the peaks. The Lorentzian width was fixed for the same component in all the measured conditions, except for the Pd(0), where asymmetry was fixed.

For the AP-GIXS measurements, a beam energy of 1240 eV, corresponding to a wavelength of 1 nm was used. The total collection solid angle was +/-12° in-plane and 24° out of plane, which translates to $q_{max}$ ~ 0.9 nm$^{-1}$ and 1.8 nm$^{-1}$, respectively. Andor iKon-L CCD mounted on the biaxial quasi-spherical manipulator was used to collect scattered photons. To minimize the influence of island formation and particle clustering, GIXS analysis was performed on a out of plane line-cut extracted at a scattering vector (q) of 0.0 nm$^{-1}$.[49] The PDDF was obtained from the inverse Fourier transform of the horizontal line-cut utilizing the BioXTAS RAW software package (see Note S18 for details on GIXS line cuts and analysis).[50–52] The electron density reconstruction was performed utilizing the BIOXTAS RAW and the integrated DENS package (details on the reconstruction are found in Note S2).[24]

We also performed AP-XPS measurements at APXPS endstation at 9.3.2 beamline of ALS Synchrotron. The nanoparticles were prepared in the same way presented for the measurements at 11.0.2. The nanoparticles were inserted into an ambient pressure chamber and heated to 700 °C under vacuum. Pd 3d and Ni 3p energy regions were measured during the heating process. After that, the nanoparticles were cooled down to room temperature where they were exposed to 1x10$^{-7}$ mbar, 1x10$^{-5}$ mbar, 1x10$^{-3}$ mbar, 1x10$^{-1}$ mbar and 5x10$^{-1}$ mbar of H$_2$. At each pressure, energy regions collected were survey, Pd 3d, Ni 3p, Pd 4p, Si 2p, C1s, and O1s. All the measurements were performed using a 695 eV photon beam energy, a pass energy of 20 eV, dwell time of 100 ms and step size of 0.1 eV and 0.5 were used for the high-resolution and survey spectra.

*Hydrogen Storage Capacity Measurements*

We determined the hydrogen storage capacity by Gas Chromatography (GC) measurements at Physics of Nanostructures Laboratory – UFRGS using a GC equipment model 310 USB from SRI Instruments. The GC equipment is equipped with a Thermal Conductivity Detector (TCD) and operates with Ar as carrier gas. A home-built reactor was used for the reactions. Initially, the $Pd_xNi_{100-x}$ nanoparticles were supported over activated carbon (Activated Charcoal Norit®) with 1 wt.% through mechanical support followed by an ultrasound batch for 20 minutes. A sample of pure activated carbon was also measured for comparison purposes. The sample was introduced in the reactor, and it was purged with Ar before inserting 1 atm Ar atmosphere. After this, the system was heated to 150 °C and remained at this temperature for 1 h to remove humidity and to clean the nanostructures surface. Then the system was purged with Ar and cooled to room temperature under 1 atm Ar atmosphere. After the thermal treatment, 1 atm $H_2$ was inserted in the reactor, and aliquots of 150 µL of the gas atmosphere inside the reactor were taken every 60 min to determine the $H_2$ amount in the inner atmosphere of the reactor by GC measurements. The comparison with the initial amount of $H_2$ enables to determine the amount of $H_2$ adsorbed by the sample. This measurement was repeated until stabilization of the $H_2$ amount inside the reactor, which occurs after around 6 h. At the end, the sample was heated to 350 °C for 1 h for $H_2$ release and thus it was exposed to the same conditions for a new measurement of $H_2$ storage.

The gravimetric capacity was calculated using:

$$gravimetric\ capacity = \frac{(nH2_{initial} - nH2_{final}) \times mH2}{mPd_xNi_{100-x} + mC}$$

where nH2 is the number of moles of $H_2$ calculated from the amount of $H_2$ present in the gas of the reactor considering it as an ideal gas and m stands for the mass. The pressure and temperature inside the reactor were constantly monitored with the use of a sensor and it remained close to 1 atm and 20 °C, but small differences observed were also considered in the determination of the gravimetric capacity.

*Scanning Electrons Microscopy (SEM) and Energy Dispersive Spectroscopy (EDS)*

We obtained the SEM images and EDS spectra at Molecular Foundry-LBL using a Zeiss Gemini Ultra-55 microscope operated with 5 kV. The samples were measured before and after the AP-XPS and AP-GIXS measurements. The SEM images were obtained by detecting secondary electrons.

**Atomic Force Microscopy (AFM)**

We conducted AFM measurements at the Imaging Facility of the Molecular Foundry-LBL using an Asylum Jupyter AFM, and a PEAKFORCE-HIRS-F-A tip from BrukerNano. The AFM was operated in tapping mode, with a drive amplitude of 100 mV and a setpoint of 60 mV. Scan speed was set to 0.75 lines per second and the resolution to 256 x 256 lines x rows.

ASSOCIATED CONTENT

**Supporting Information**.
TEM, XPS, EXAFS, EDS, Gravimetric capacity, AFM, GIXS and EELS data and analysis.

AUTHOR INFORMATION

**Corresponding Author**

*Corresponding authors. Email: SNemsak@lbl.gov; bernardi@if.ufrgs.br;

**Author Contributions**

The manuscript was written through contributions of all authors. All authors have given approval to the final version of the manuscript.

ACKNOWLEDGMENT

This research used APPEXS endstation at beamlines 9.3.2 and 11.0.2.1 of the Advanced Light Source, which is a DOE Office of Science User Facility under contract no. DE-AC02-05CH11231. The authors thank ALBA Synchrotron staff for the measurements performed at NOTOS (BL-16) beamline (proposal 2021095397). The authors also thank CNANO-UFRGS and CM-UFMG staff for their assistance. This study was funded by FAPERGS (19/2551-0001752-9, 19/2551-0000703-5, and 23/2551-0000177-2), CNPq (310142/2021-0) and CAPES (Finance Code 001). LPM, AST and FB thank the CNPq for the research grant. LPM thanks CAPES for the research grant. MJ is funded by the Catalysis program FWP CH030201. CE thanks the MICINN-FEDER funding through the PID2021-124572OB-C33 grant.  Work at the Molecular Foundry was supported by the Office of Science, Office of Basic Energy Sciences, of the U.S. Department of Energy under Contract No. DE-AC02-05CH11231.

REFERENCES

(1) Karpa, W.; Grginović, A. (Not so) Stranded: The Case of Coal in Poland. *Energies* **2021**, *14*, 8476.

(2) Mazloomi, K.; Gomes, C. Hydrogen as an Energy Carrier: Prospects and Challenges. *Renew. Sustain. Energy Rev.* **2012**, *16*, 3024–3033.

(3) DOE Technical Targets for Onboard Hydrogen Storage for Light-Duty Vehicles. https://www.energy.gov/eere/fuelcells/doe-technical-targets-onboard-hydrogen-storage-light-duty-vehicles (accessed February 21st, 2025), 2017.

(4) Usman, M. R. Hydrogen Storage Methods: Review and Current Status. *Renew. Sustain. Energy Rev.* **2022**, *167*, 112743.

(5) Jena, P. Materials for Hydrogen Storage: Past, Present, and Future. *J. Phys. Chem. Lett.* **2011**, *2*, 206–211.

(6) Konda, S. K.; Chen, A. Palladium Based Nanomaterials for Enhanced Hydrogen Spillover and Storage. *Mater. Today (Kidlington)* **2016**, *19*, 100–108.

(7) Chen, B. W. J.; Mavrikakis, M. Effects of Composition and Morphology on the Hydrogen Storage Properties of Transition Metal Hydrides: Insights from PtPd Nanoclusters. *Nano Energy* **2019**, *63*, 103858.

(8) Thill, A. S.; Lima, D. S.; Perez-Lopez, O. W.; Manfro, R. L.; Souza, M. M. V. M.; Baptista, D. L.; Archanjo, B. S.; Poletto, F.; Bernardi, F. A new non-complex synthesis of NiO nanofoams for hydrogen storage applications. *Mater. Adv.* **2023**, *4,* 476-480.

(9) Ren, J.; Liao, S.; Liu, J. Hydrogen Storage of Multiwalled Carbon Nanotubes Coated with Pd-Ni Nanoparticles under Moderate Conditions. *Chin. Sci. Bull.* **2006**, *51*, 2959–2963.

(10) Sano, N.; Taniguchi, K.; Tamon, H. Hydrogen Storage in Porous Single-Walled Carbon Nanohorns Dispersed with Pd–Ni Alloy Nanoparticles. *J. Phys. Chem. C Nanomater. Interfaces* **2014**, *118*, 3402–3408.

(11) Campesi, R.; Cuevas, F.; Leroy, E.; Hirscher, M.; Gadiou, R.; Vix-Guterl, C.; Latroche, M. In Situ Synthesis and Hydrogen Storage Properties of PdNi Alloy Nanoparticles in an Ordered Mesoporous Carbon Template. *Microporous Mesoporous Mater.* **2009**, *117*, 511–514.

(12) Militello, M. C.; Simko, S. J. Palladium Oxide (PdO) by XPS. *Surf. Sci. Spectra* **1994**, *3*, 395–401.

(13) Chen, L. F.; Wang, J. A.; Valenzuela, M. A.; Bokhimi, X.; Acosta, D. R.; Novaro, O. Hydrogen Spillover and Structural Defects in a PdO/Zirconia Nanophase Synthesized through a Surfactant-Templated Route. *J. Alloys Compd.* **2006**, *417*, 220–223.

(14) Figueiredo, W. T.; Prakash, R.; Vieira, C. G.; Lima, D. S.; Carvalho, V. E.; Soares, E. A.; Buchner, S.; Raschke, H.; Perez-Lopez, O. W.; Baptista, D. L.; Hergenröder, R.; Segala, M.; Bernardi, F. New Insights on the Electronic Factor of the SMSI Effect in Pd/TiO$_2$ Nanoparticles. *Appl. Surf. Sci.* **2022**, *574*, 151647.

(15) Veser, G.; Wright, A.; Caretta, R. On the oxidation–reduction kinetics of palladium. *Catal. Letters* **1999**, *58*, 199-206.

(16) Bird, R. J.; Swift, P. Energy Calibration in Electron Spectroscopy and the Re-Determination of Some Reference Electron Binding Energies. *J. Electron Spectros. Relat. Phenomena* **1980**, *21*, 227–240.

(17) Mejías, J. A.; Jiménez, V. M.; Lassaletta, G.; Fernández, A.; Espinós, J. P.; A. R. González-Elipe*. Interpretation of the Binding Energy and Auger Parameter Shifts Found by XPS for TiO$_2$ Supported on Different Surfaces. *J. Phys. Chem.* **1996**, *100*, 16255–16262.

(18) Mansour, A. N. Characterization of NiO by XPS. *Surf. Sci. Spectra* **1994**, *3*, 231–238.

(19) Bennett, P. A.; Fuggle, J. C. Electronic Structure and Surface Kinetics of Palladium Hydride Studied with X-Ray Photoelectron Spectroscopy and Electron-Energy-Loss Spectroscopy. *Phys. Rev. B Condens. Matter* **1982**, *26*, 6030–6039.

(20) Yeh, J. J.; Lindau, I. Atomic Subshell Photoionization Cross Sections and Asymmetry Parameters: $1 \leqslant Z \leqslant 103$. At. Data Nucl. *Data Tables* **1985**, *32*, 1–155.

(21) Schwartzkopf, M.; Buffet, A.; Körstgens, V.; Metwalli, E.; Schlage, K.; Benecke, G.; Perlich, J.; Rawolle, M.; Rothkirch, A.; Heidmann, B.; Herzog, G.; Müller-Buschbaum, P.; Röhlsberger, R.; Gehrke, R.; Stribeck, N.; Roth, S. V. From Atoms to Layers: In Situ Gold Cluster Growth Kinetics during Sputter Deposition. *Nanoscale* **2013**, *5*, 5053.

(22) Guinier, G. *Small-Angle Scattering of X-Rays*; John Wiley & Sons: New York, 1955.

(23) Schnablegger, H.; Singh, Y. *The SAXS Guide*; Anton Paar GmbH: Graz, 2023.

(24) Grant, T. D. Ab Initio Electron Density Determination Directly from Solution Scattering Data. *Nat. Methods* **2018**, *15*, 191–193.

(25) Anitas, E. M. Small-Angle Scattering from Fractals: Differentiating between Various Types of Structures. *Symmetry (Basel)* **2020**, *12*, 65.

(26) Ruban, A. V.; Skriver, H. L.; Nørskov, J. K. Surface Segregation Energies in Transition-Metal Alloys. *Phys. Rev. B Condens. Matter* **1999**, *59*, 15990–16000.

(27) Matte, L. P.; Kilian, A. S.; Luza, L.; Alves, M. C. M.; Morais, J.; Baptista, D. L.; Dupont, J.; Bernardi, F. Influence of the $CeO_2$ Support on the Reduction Properties of $Cu/CeO_2$ and $Ni/CeO_2$ Nanoparticles. *J. Phys. Chem. C Nanomater. Interfaces* **2015**, *119*, 26459–26470.

(28) Mavrikakis, M.; Hammer, B.; Nørskov, J. K. Effect of Strain on the Reactivity of Metal Surfaces. *Phys. Rev. Lett.* **1998**, *81*, 2819–2822.

(29) Tew, M. W.; Miller, J. T.; van Bokhoven, J. A. Particle Size Effect of Hydride Formation and Surface Hydrogen Adsorption of Nanosized Palladium Catalysts: $L_3$ Edge vs K Edge X-Ray Absorption Spectroscopy. *J. Phys. Chem. C Nanomater. Interfaces* **2009**, *113*, 15140–15147.

(30) Oumellal, Y.; Provost, K.; Ghimbeu, C. M.; de Yuso, A. M.; Zlotea, C. Composition and Size Dependence of Hydrogen Interaction with Carbon Supported Bulk-Immiscible Pd-Rh Nanoalloys. *Nanotechnology* **2016**, *27*, 465401.

(31) Matte, L. P.; Khan, W.; Thill, A. S.; Escudero, C.; Poletto, F.; Bernardi, F. Controllable Morphology of Pd Nanostructures: From Nanoparticles to Nanofoams. *Mater. Res. Express* **2024**, *11*, 105010.

(32) Liao, H.; Fisher, A.; Xu, Z. J. Surface Segregation in Bimetallic Nanoparticles: A Critical Issue in Electrocatalyst Engineering. *Small* **2015**, *11*, 3221–3246.

(33) Jain, A.; Ong, S. P.; Hautier, G.; Chen, W.; Richards, W. D.; Dacek, S.; Cholia, S.; Gunter, D.; Skinner, D.; Ceder, G.; Persson, K. A. Commentary: The Materials Project: A Materials Genome Approach to Accelerating Materials Innovation. *APL Mater.* **2013**, *1*, 011002.

(34) Ong, S. P.; Wang, L.; Kang, B.; Ceder, G. Li−Fe−P−$O_2$ Phase Diagram from First Principles Calculations. *Chem. Mater.* **2008**, *20*, 1798–1807.

(35) Aykol, M.; Dwaraknath, S. S.; Sun, W.; Persson, K. A. Thermodynamic Limit for Synthesis of Metastable Inorganic Materials. *Sci. Adv.* **2018**, *4*, eaaq0148.

(36) Bartel, C. J.; Millican, S. L.; Deml, A. M.; Rumptz, J. R.; Tumas, W.; Weimer, A. W.; Lany, S.; Stevanović, V.; Musgrave, C. B.; Holder, A. M. Physical Descriptor for the Gibbs Energy of Inorganic Crystalline Solids and Temperature-Dependent Materials Chemistry. *Nat. Commun.* **2018**, *9*, 4168.

(37) van de Walle, A.; Ceder, G. The Effect of Lattice Vibrations on Substitutional Alloy Thermodynamics. *Rev. Mod. Phys.* **2002**, *74*, 11–45.

(38) Kapoor, V.; Singh, B.; Sai Gautam, G.; Cheetham, A. K.; Canepa, P. Rational Design of Mixed Polyanion Electrodes $Na_xV_2P_{3-i}(Si/S)_iO_{12}$ for Sodium Batteries. *Chem. Mater.* **2022**, *34*, 3373–3382.

(39) Tayal, A.; Seo, O.; Kim, J.; Kobayashi, H.; Yamamoto, T.; Matsumura, S.; Kitagawa, H.; Sakata, O. Mechanism of Hydrogen Storage and Structural Transformation in Bimetallic Pd–Pt Nanoparticles. *ACS Appl. Mater. Interfaces* **2021**, *13*, 23502–23512.

(40) Yamauchi, M.; Kobayashi, H.; Kitagawa, H. Hydrogen Storage Mediated by Pd and Pt Nanoparticles. *Chemphyschem* **2009**, *10*, 2566–2576.

(41) Bergerhoff, G.; Berndt, M. Inorganic Structural Chemistry with Crystallographic Databases. *Acta Crystallogr. A* **1996**, *52*, C314–C314.

(42) Rodríguez-Carvajal, J. Recent Advances in Magnetic Structure Determination by Neutron Powder Diffraction. *Physica B Condens. Matter* **1993**, *192*, 55–69.

(43) Thompson, P.; Cox, D. E.; Hastings, J. B. Rietveld Refinement of Debye–Scherrer Synchrotron X-Ray Data from $Al_2O_3$. *J. Appl. Crystallogr.* **1987**, *20*, 79–83.

(44) Koningsberger, D. C.; Mojet, B. L.; van Dorssen, G. E.; Ramaker, D. E. Fundamental Principles and Data Analysis. *Top. Catal.* **2000**, *10*, 143–155.

(45) Ravel, B.; Newville, M. *ATHENA,ARTEMIS,HEPHAESTUS*: Data Analysis for X-Ray Absorption Spectroscopy Using *IFEFFIT*. *J. Synchrotron Radiat.* **2005**, *12*, 537–541.

(46) Kersell, H.; Chen, P.; Martins, H.; Lu, Q.; Brausse, F.; Liu, B.-H.; Blum, M.; Roy, S.; Rude, B.; Kilcoyne, A.; Bluhm, H.; Nemšák, S. Simultaneous Ambient Pressure X-Ray Photoelectron Spectroscopy and Grazing Incidence x-Ray Scattering in Gas Environments. *Rev. Sci. Instrum.* **2021**, *92*, 044102.

(47) Pillo, T.; Zimmermann, R.; Steiner, P.; Hüfner, S. The Electronic Structure of PdO Found by Photoemission (UPS and XPS) and Inverse Photoemission (BIS). *J. Phys. Condens. Matter* **1997**, *9*, 3987–3999.

(48) Militello, M. C.; Simko, S. J. Elemental Palladium by XPS. *Surf. Sci. Spectra* **1994**, *3*, 387–394.

(49) Jaugstetter, M.; Qi, X.; Chan, E. M.; Salmeron, M.; Wilson, K. R.; Nemšák, S.; Bluhm H. Direct Observation of Morphological and Chemical Changes during the Oxidation of Model Inorganic Ligand-Capped Particles. *ACS Nano* **2025** *19*, 418-426

(50) Putnam, C. D.; Hammel, M.; Hura, G. L.; Tainer, J. A. X-Ray Solution Scattering (SAXS) Combined with Crystallography and Computation: Defining Accurate Macromolecular Structures, Conformations and Assemblies in Solution. *Q. Rev. Biophys.* **2007**, *40*, 191–285.

(51) Jacques, D. A.; Trewhella, J. Small-angle Scattering for Structural Biology—Expanding the Frontier While Avoiding the Pitfalls. *Protein Sci.* **2010**, *19*, 642–657.

(52) Feigin, L. A.; Svergun, D. I. *Structure Analysis by Small-Angle X-Ray and Neutron Scattering*; Taylor, G. W., Ed; Springer Nature: New York, 1987.

TABLE OF CONTENT (TOC)

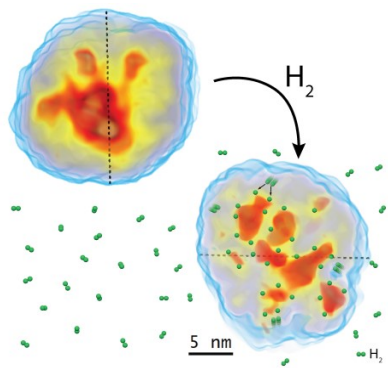